\documentstyle{l-aa}

\newcommand{\mathrm}{\rm}
\newcommand{\D}{\mathrm d}
\newcommand{\e}{\mathrm e}

\newcommand{\I}[1]{_{\mathrm #1}}
\newcommand{\Mi}{M\I{i}}
\newcommand{\Zi}{Z\I{i}}
\newcommand{\ea}{et al.}
\newcommand{\Lsun}{\mbox{$L_{\odot}$}}
\newcommand{\Msun}{\mbox{$M_{\odot}$}}
\newcommand{\q}{$\!\!\!$}               
\newcommand{\tip}{\mbox{$\tau_{\mathrm ip}$}}
\newcommand{\LA}{\mbox{$L_{\mathrm A}$}}
\newcommand{\LB}{\mbox{$L_{\mathrm B}$}}
\newcommand{\LC}{\mbox{$L_{\mathrm C}$}}
\newcommand{\LD}{\mbox{$L_{\mathrm D}$}}
\newcommand{\Mc}{M_{\mathrm c}}
\newcommand{\Mci}{M_{\mathrm c,0}}

\newcommand{\DMc}{\Delta M_{\mathrm c}}
\newcommand{\Mm}{M_{\mathrm m}}
\newcommand{\beq}[1]{\vspace{-0.5mm} \begin{equation} \label{#1}}
\newcommand{\beqa}{\begin{eqnarray}}
\newcommand{\eeq}{\end{equation}}\vspace{-0.2mm}
\newcommand{\eeqa}{\end{eqnarray}}
\renewcommand{\labelitemi}{-}

\newcounter{subeqn}
\newcommand{\bsubeqa}[2]{\beqa\label{#1}\nonumber\eeqa\vspace*{-12.5mm}\renewcommand{\theequation}{\arabic{equation}\alph{subeqn}}\setcounter{subeqn}{1}\beqa\label{#2}}
\newcommand{\nsubeqn}[1]{\addtocounter{subeqn}{1}\addtocounter{equation}{-1}\label{#1}}
\newcommand{\esubeqa}{\eeqa\renewcommand{\theequation}{\arabic{equation}}}

\include{psfig}

\begin{document}


\thesaurus{06(03.13.4; 08.16.4; 08.05.3; 08.06.3)}

\title{New input data for synthetic AGB evolution}

\author{J.~Wagenhuber and M.A.T. Groenewegen}

\institute{Max-Planck-Institut f\"{u}r Astrophysik,
        Karl-Schwarzschild-Stra{\ss}e~1,
        D-85740 Garching, Germany}
\offprints{M.A.T. Groenewegen. E-mail: groen@mpa-garching.mpg.de}
\date{Received ; accepted }
\maketitle

\begin{abstract}

Analytic formulae are presented to construct detail\-ed secular
lightcurves of both early asymptotic giant branch (AGB) and thermally
pulsing AGB stars. They are based on an extensive grid of evolutionary
calculations, performed with an updated stellar evolution code.
Basic input parameters are the initial mass $\Mi$, $0.8 \le
\Mi/M_{\odot} \le 7$, metallicity $\Zi =0.0001, 0.008, 0.02$, and the
mixing length theory (MLT) parameter. The formulae allow for two
important effects, namely that the first pulses do not reach the full
amplitude, and hot bottom burning (HBB) in massive stars, which are
both not accounted for by \mbox{core mass -} luminosity relations of
the usual type.

Furthermore, the dependence of the effective temperature and a few
other quantities characterizing the conditions at the base of the
convective envelope, which are relevant for HBB, are investigated as
functions of luminosity, total and core mass for different
formulations of the convection theory applied, MLT or Canuto \&
Mazzitelli's (\cite{can:maz}) theory.



\end{abstract}

\section{Introduction}
\label{sec-intro}

As a matter of fact, full AGB stellar evolution calculations are
unable to provide the statistical information needed for purposes like
population synthesis.
There are three main reasons for this.

First, the calculations are rather lengthy, reflecting the complexity
of the inner structure of AGB stars and of the temporal evolution due
to thermal pulses (TPs, also called helium shell flashes; see e.\ g.\
Iben \& Renzini \cite{iben:renz}; Boothroyd \& Sackmann
\cite{boo:sac}, in the following BS88; Lattanzio \cite{laz86},
LA86; Wagenhuber \& Weiss \cite{wag:wei:a},WW94; and Bl\"ocker
\cite{bloe:a}, B95).


Second, the calculations involve at least two poorly known parameters,
the mass loss efficiency $\eta$ and the mass loss prescription, and
the mixing length (MLT) parameter $\alpha$, each of them only
meaningful in the framework of the {\it a priori} chosen description
of the mass loss and convection theory, respectively. Since it has
become clear that AGB evolution is dominated by mass loss
(Sch\"onberner \cite{schb:a}), numerous formalisms giving the mass
loss rate as a function of the stellar parameters have been applied
(e.g. Reimers 1975, Iben \& Renzini \cite{iben:renz}, Mazzitelli \&
D'Antona \cite{mazz:dant}, Wood \& Faulkner \cite{woo:fau},
Vassiliadis \& Wood \cite{vw93} VW93, B95).  The uncertainties were
cast into the ``efficiency'' parameter, which depends on the mass loss
prescription used, and which is actually a parameter that can be
calibrated from AGB population synthesis models (e.g. GdJ, Groenewegen
\& de Jong 1994).

In addition, the MLT parameter is related to a series of problems
concerning (a) the correct effective temperature scale of late-type
stars derived from observations,
(b) the opacity at low temperatures (Alexander \& Ferguson
\cite{alex:ferg}), (c) the time-dependent problem of dust formation
(Fleischer et al. \cite{flei:gau}, and references
therein), and (d) the theory of convection itself.

The third restriction of the use of full evolutionary calculations is
the occurrence of the third dredge-up as a consequence of TPs, which
is considered to be responsible for the formation of carbon stars.
However, canonical stellar evolution calculations do not consistently
predict the formation of carbon stars in the mass range that they are
observed, roughly between 1.2-1.5 \Msun\ depending on metallicity and
about 5 \Msun\ (e.g. GdJ, Groenewegen et al. 1995).
The explanation is still under debate and recent progress has 
been made (see e.\ g.\ 
Frost \& Lattanzio \cite{frost}, Staniero et al. 1997, Herwig et al. 1997) 
 but not to the degree that the results can easily be included in
synthetic AGB models. Therefore these models introduce a fudge
parameter to describe the third dredge-up, in particular the
dredge-up efficiency $\lambda$.

To summarize: In order to describe the evolution of whole
stellar populations using synthetic calculations, one has to explore a
parameter space with at least five dimensions: $(\Mi, \Zi, \alpha,
\eta, \lambda)$. This has been done in the past e.\ g.\ by Renzini \&
Voli (\cite{renz:voli}), J\o rgensen (\cite{uffe}), Groenewegen \& de
Jong (\cite{gro:dej}, GdJ) and Marigo \ea\
(\cite{marigo}). Although these works have contributed a lot to an
improved understanding of AGB evolution, there are some obvious 
shortcomings: e.g. the data collected by GdJ and frequently used since then
are partially incomplete, making some {\it ad hoc} inter- and
extrapolations necessary, since results obtained in
three decades are combined with each other,
and they are oversimplified in some respects, e. g. concerning hot
bottom burning (HBB).
Furthermore, some data such as the rapid luminosity variations during a
TP were neglected by many authors, although they are needed for some
of the applications mentioned below.

The aim of this paper is to provide essential theoretical data for
applications like classical synthetic AGB evolution. In this sense
this paper provides a fully updated and improved set of relations with
respect to GdJ.


The paper is organized as follows.  In the next section some
terminology is introduced, and the intrinsic errors due to physical
assumptions are discussed.  After this, the full evolution
calculations up to and along the early (E-)AGB are shortly described
(Sect. 3).  The fourth section contains the recipes necessary to
construct the secular lightcurves on the TP-AGB as a function of time
for all relevant initial total masses and metallicities. Finally,
properties of the effective temperature and the conditions at the
bottom of the convective envelope for various model assumptions are
discussed in Sect. 5.


\section{General remarks and computational details}
\label{sec-gr}

\subsection{Some definitions}
\label{ssec-defs}

Let us first define some quantities and index labels that
are used throughout the rest of this paper.

A {\it thermal pulse cycle} (TPC) is the time interval from a
local maximum of helium burning, through quiescent hydrogen burning,
up to the next TP. All quantities that refer to a TPC as a whole are
defined to be functions of the conditions at the beginning of the
TPC. The {\it first TP} is the one in which the maximum (integrated)
luminosity produced by He-burning exceeds the maximum H-burning
luminosity prior to this for the first time. Previously there may be
{\it ``pre-pulses''}.

{\it Uppercase} letters denote quantities that depend  {\it
linearly} on parameters. Time intervals
(in years) are denoted by $\tau$, in particular the duration of a TPC,
the {\it interpulse time}, by \tip. Luminosity and mass $(L, M)$ are
given in solar units as usual, temperatures $T$ in K, the abundances
of hydrogen $X$, helium $Y$ and metals $Z$ in relative mass units
($X+Y+Z=1$), and all others in cgs units.

{\it Lowercase} letters are used for decadic logarithms of the above
quantities, e. g. $l \equiv \lg L/\Lsun$, except for $z \equiv \lg
Z/0.02$ being the logarithmic metallicity scaled to solar.

The {\it core} is defined to be the part of a star inside the location
where the local hydrogen content reaches half the photospheric
value. Since the H-burning shell is extremely narrow (both in radius
and relative mass units), this essentially coincides with all other
definitions of the core mass in the literature. The {\it core growth}
$\DMc$ is defined as $\int \max \{ \frac{\D \Mc}{\D \tau}, 0 \} \D
\tau$. Usually, $\DMc = \Mc(\tau) - \Mci$, where $\Mci$ is the core
mass at the first TP.

The {\it mantle}
is the part between the photosphere and the core. In terms of
the relative mass content, it is almost identical with what is
frequently called ``(convective) envelope''.

One has to distinguish three different TP-AGB phases, defined in
WW94. There are:

\begin{itemize}
\renewcommand{\labelitemi}{(i)}
\item The {\it transition phase} from the E-AGB to the first TP, in
which pre-pulses may take place.
\renewcommand{\labelitemi}{(ii)}
\item The {\it turn-on phase}, comprising about the first ten TPs.
All global quantities are aiming at their asymptotic behaviour, but
still the deviations are significant (up to 60\%). This is called {\it
turn-on effect} or  TOE here.
\renewcommand{\labelitemi}{(iii)}
\item The {\it asymptotic phase}, where the global quantities have
reached their asymptotic behaviour.
\end{itemize}

The MLT parameter $\alpha$ may be defined either for the MLT or Canuto
\& Mazzitelli's (\cite{can:maz}) theory (CMT) and is labelled
accordingly, if necessary.

The three sets of calculations for $\Zi = 0.02, 0.008$ and $10^{-4}$
are called {\it pop I, LMC} and {\it pop II} in the following,
respectively.

\vspace{2mm}
\noindent{\bf Subscripts:}
\vspace{-7mm}

\begin{figure}
 \centerline{\psfig{figure=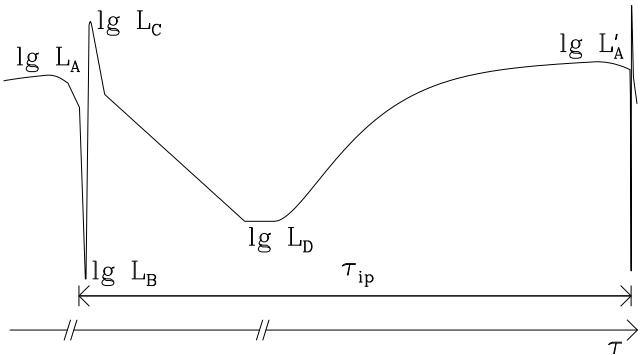,width=8.5cm}}
\caption[]{A typical TPC as produced by the code described in the
text. The labels illustrate some of the definitions in sect.\
\protect\ref{ssec-defs}. The time axis has been stretched around the
first TP shown.}
  \label{fig-scetch}
\end{figure}

\begin{tabbing}
\hspace*{1.5em} \= \\
{\bf b}:   \> 	Quantities defined at the bottom of the convective mantle,\\
{\bf c}:   \>	concerning the core and\\
{\bf m}:   \>	the mantle.
\end{tabbing}\vspace{-10mm}
\begin{tabbing}
\hspace*{1.5em} \= \\
{\bf A},{\bf B},{\bf C},{\bf D}: Denote luminosity extrema in the TPC
(Fig. \ref{fig-scetch}):\\
{\bf A} -  \>	the {\it ``slow maximum''} during quiescent H-burning,\\
{\bf B} -  \>	the {\it ``rapid dip''} following a TP,\\
{\bf C} -  \>	the {\it ``rapid peak''} after this,\\
{\bf D} -  \>	the {\it ``slow dip''} at the transition from He- to
H-burning.
\end{tabbing}\vspace{-10mm}
\begin{tabbing}
\hspace*{1.5em} \= \\
{\bf i}:   \>	Initial values on the ZAMS, and\\
{\bf f}:   \>	final values after the TP-AGB phase.\\
{\bf H}:   \>	Something produced by or related to hydrogen or\\
{\bf He}:  \>	helium burning.\\
{\bf 0}:   \>	Quantities at the beginning of an evolutionary phase\\
	   \>	as a whole, like the E-AGB and the TP-AGB,\\
{\bf TP}:  \>	at the beginning of a TPC, and\\
{\bf 1}:   \>	at the end of the corresponding phase.\\
{\bf *}:   \>	Denotes ``effective'' values measured at the photosphere.\\
\end{tabbing}\vspace{-6mm}

\subsection{Intrinsic errors and physical assumptions}
\label{ssec-err}

About 40 model sequences were run, with varying initial mass,
metallicity or physical input parameters. Initial masses are between
0.8 $\le \Mi \le$ 7, the metallicity is $\Zi$ = 0.02, 0.008, 0.0001.
The standard values for $Y$ on the main sequence were 0.28, 0.2651 and
0.25 for pop~I, LMC and pop~II, respectively.
The metal mix is according to Anders \& Grevesse (\cite{an:gr}) for
pop~I, scaled by the appropriate factor for pop~II.  For the LMC
mixture, the abundances of C,N and O, which are considered explicitly in
the equation of state and the nuclear reactions, deviate from the
solar values following Russell \& Bessell (\cite{rus:bes}) and
Russell \& Dopita (\cite{rus:dop}). Opacity tables were always used,
according to their X,Y and Z values, neglecting the internal metal ratios.



During the full stellar evolution model calculations new input physics
became available which allowed us to test for any differences indiced
by them.
Two different physical models were used for the
pop~I calculations and $\Mi \in \{1, 1.5, 2, 3, 5, 7\}$: one with
nuclear equilibrium rates and opacities from the collection of older
opacity tables by Weiss et al. (\cite{WKM}), as explained in
WW94, and the other one with a nuclear network, and more recent
opacities taken from Rogers \& Iglesias (\cite{OPAL}), in the
following referred to as models A and B, respectively. For $T <
10^4$K, the OPAL tables were completed by LAOL data. Both sets agree
well around $T \approx 10^4$K. The LMC calculations were carried out
with model A, model B was employed for pop~II. Our nuclear network
comprises the following species relevant for the AGB: $^1$H, $^3$He,
$^4$He, $^{12}$C, $^{13}$C, $^{14}$N, $^{15}$N, $^{16}$O, $^{17}$O,
$^{20}$Ne, $^{24}$Mg. A leakage out of the CNO cycle, which is unimportant
for the energy generation rate (Mowlavi et al. 
\cite{mow:jor}), was not taken into account. For more details, see
Weiss \& Truran (\cite{wei:trur}), Weiss et al. 
(\cite{wwd}) and Wagenhuber (\cite{phd}). The models were evolved off
the ZAMS through the core helium flash
(for $\Mi \le 2$) up to the AGB without manual interventions and with
a high numerical resolution (see WW94).

When comparing model A and B calculations, one finds as the major
effect that for $\Mi \ga 4$ the initial core mass $\Mci$ as a function
of $\Mi$ increases slightly faster for model B than for model A (see
Fig. \ref{fig-Mc0}). The greatest difference of $0.08\Msun$ occurs for
$\Mi = 7$ (the OPAL data for $\Mci$ agree with those of D'Antona \&
Mazzitelli (\cite{dan:maz}, DAM96) to within $0.01\Msun$). Accordingly, 
the massive stars in model B reach somewhat higher luminosities than
their counterparts in model A with the same $\Mi$.
However, on the TP-AGB luminosities and time scales, {\it when
considered as functions of} $\Mc$ {\it and} $\DMc$, agree with each
other to within 12\%, the main contribution being that the TOE tends
to be stronger for model B.

The convective core of the Horizontal Branch models grow slowly into
the semi-converctive regions, which, due to the use of the
Schwarzschild-criterion, are formally stable stable in our
solutions. The growth is due to the smoothing of (unphysical)
molecular weight steps in the grid-control routine of the
code. However, in the B-sequences, the growth happened in discrete
``breathing pulses'' (Castellani \ea\ \cite{ccpt}). The size and
occurence depends on both physical (mixture, opacities) and numerical
details. For increased resolution in both space and time their extent
is reduced and the cores grow in a more continous way.



The standard choice for the MLT parameter was $\alpha = 1.5$, which
together with the adopted opacities for low temperatures yielded
somewhat lower $T_*$ (i.e. the effective temperature following our
nomenclature) when compared to results by DAM96 and B95. The
code used in the present work needs a MLT parameter approximately
0.25 and 0.2 larger, respectively, to reproduce their results.  In
order to fit the present sun with OPAL, $\alpha = 1.578$ and $Y =
0.2803$ is needed, very close to the standard values of 1.50 and 0.28,
giving rise to a possible shift of $T_*$ by far within the
uncertainties of the observed AGB effective temperature scale.  Since
the dependence of all results on $Z$ is the dominating one,
uncertainties due to the variations of $Y$ for a given $Z$ of about
$\pm 0.03$ are neglected.

Additionally, some TPCs were repeated with the CMT formulation of
convection instead of standard MLT. This was done with a mixing length
$\Lambda$ proportional to the local pressure scale heigth $H\I{p}$,
i. e. $\Lambda = \alpha\I{CMT} \, H\I{p}$. The choice $\alpha\I{CMT} =
0.65$ reproduced the effective temperatures obtained with MLT and
$\alpha\I{MLT} = 1.5$ very well (see Sect. \ref{ssec-teff}). Beyond
this, time-dependence according to the model by Arnett (\cite{arnett})
and a turbulent pressure model according to CMT were invoked in some
test sequences.  These tests reveal that the various convection
formalisms, too, do not influence luminositiy and timescales  
by more than 2\%, except if
HBB is operating (see Sect. \ref{ssec-LA}).  The time dependent
convection models give rise to two interesting effects during the
``rapid luminosity peak'': $T_*$ is cooler by about 100K than during
the quiescent phases, and $T\I{b}$ of low-mass stars, which usually
never exceeds $1.5\times 10^6$ K, reaches peak values of $5\times 10^6$K
shortly after a TP, since the bottom of the mantle convection zone
cannot retreat as quickly as in the case of ``instantaneous''
convection. However, $L_*$ and the TP timescales are unaffected.

\begin{figure}[t]
 \centerline{\psfig{figure=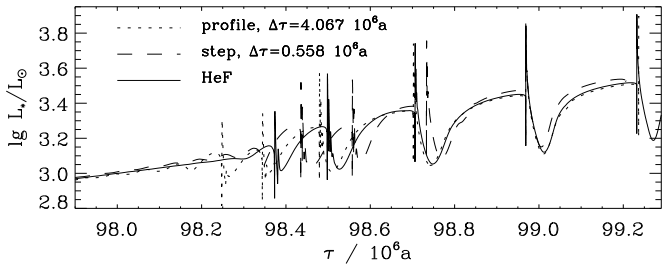,width=8.7cm}}
\caption[]{The lightcurves during the first TPCs for the full
 evolution through the core helium flash (full line) and two cases of
 artificial initial models (see text). The time axis (in Myr since the
 onset of core helium burning) of the latter two has been shifted by
 the amount indicated so that the last but one TPs lie on top of each
 other.}  \label{fig-HB-LLL}
\end{figure}

These calculations furthermore allow to estimate the impact of the way
the initial models were generated on the TP-AGB. Instead of full
calculations through the core helium flash, often the chemical profile
of the last model at the tip of the RGB (red giant branch), or an
artificial step profile, are used to obtain a core helium burning
model by means of an explicit integration method. These three
possibilities have been carried out for a pop~II, $\Mi = 1.25\Msun$
model.  The luminosities at the onset of the core helium burning
phases, which last for 62, 56 and 78 Myr, are $79.8, 81.1$ and
$82.3\Lsun$ for the full evolution, H-profile and step profile cases,
respectively. Only the latter experiences a breathing pulse.  The
beginning of the TP-AGB is shown in Fig. \ref{fig-HB-LLL}. The
sequences evolved off artificial initial models show an irregular
behaviour during the first TPCs. This leads to the conclusion that the
composition profile of the H-shell provides a rather long-lasting
memory, which may change $L_*$ and \tip\ at the first TP by about
10\%, and will therefore mildly influence the evolution of low-mass
stars, since these experience only few TPs (GdJ).

\subsection{Outline of the method}
\label{ssec-meth}

Here we briefly outline the method used to derive analytical relations
from the full calculations, using the core mass-luminosity relation as
an example. 

It has been known for a long time that two important quantities
characterizing the TP-AGB, \LA\ and $\lg \tip$, approximately are
linear functions of the core mass (Paczy\'{n}ski \cite{pac:a},
\cite{pac:b}). However, many different formulae were
published (e.g. Iben \& Truran \cite{ib:tr} (IT78), Wood \& Zarro
\cite{wo:za} (WZ81), LA86, BS88).


To find an appropriate description of our data, first $L$ and $\lg
\tau$ were assumed to depend linearly on $\Mc$, and each of the three
sets of calculations (pop~I, II and LMC) was considered separately
(sometimes a quadratic dependance was found to be more appropriate).
A random subset of data from TPCs in the asymptotic regime without
HBB, distributed within $0.55 < \Mc < 0.80$, was used to obtain a
rough estimate for the coefficients, and to determine if the latter
are constant, or depend linearly on $\Zi$ or $\lg \Zi$.  One thus
obtains a first guess relation $g^{(1)}(\Mc, \Zi)$.

Next, the residua due to the TOE are considered, still ignoring HBB
for the moment.  They can well be described by an exponentially
decaying function of the core growth, i. e. the ansatz \mbox{$g^{(2)}
\approx \lg |f - g^{(1)}|$} is a linear function of $\DMc$, with
coefficients depending on $\Mc$ or $\Mci$ and $\Zi$ or $\lg \Zi$,
which are determined by a non-linear fit for a subset of
all the data. To continue, the least significant terms in the new
residual function
are removed. The resulting formula is checked against the full
available data set. One finds that the data points previously not
taken into account are well predicted, with deviations of the order of
less than twice the mean error with respect to the subset data.

The effects of HBB are strongly non-linear functions of the total mass
and the MLT parameter $\alpha$. One calculation with $\alpha = 2$
($\Mi = 5\Msun$) was carried out, and data from DAM96 and B95 were
additionally considered. A large number of analytic forms with the
correct qualitative behaviour were tested.

In a final step, all coefficients are improved simultaneously by
applying a non-linear fitting procedure to the complete data set.
The values of the parameters are given with as many digits as
necessary to obtain deviations of less than 1\% with respect to
results calculated with 16 digits. The number of digits must not be
mistaken as an indication that the parameters are ``known'' to this
precision, since changing one parameter usually could be compensated
by changes of another one, at the expense that some data points were
reproduced considerably worse.

%
%


\subsection{Limits of applicability}
\label{ssec-apli}

All the formulae presented below should be considered as interpolations,
valid for the parameter space $0.8 \la \Mi \la 7$ and $10^{-4} \le \Zi
\la 0.02$. 
Below $\Zi = 10^{-4}$, new physical effects occur (see Cassisi et
al. \cite{cas2:tor}) that partially invalidate the descriptions given below.

With the standard choice of the MLT-parameter, $\alpha = 1.5$, almost
no HBB takes place, and the resulting $T_*$ are close to the lower
limit of the what is allowed by observations. Hence $\alpha < 1.4$ is
meaningless. On the other hand, no data with $\alpha > 2.8$ are
available, so that larger values should be avoided too.


\section{The evolution prior to the first thermal pulse}
\label{sec-RGB}

In synthetic AGB evolution calculations, the evolution prior to the
first TP must be taken into account to determine the total mass left
at the onset of TPs.

\subsection{Up to the core helium flash}
\label{ssec-rgb}

Low-mass stars that experience the core helium flash may lose a
significant amount of mass on the first giant branch. Jimenez \&
MacDonald (\cite{jim:mcd}) give a core mass-luminosity relation (CMLR)
for the RGB, which in terms of the variables introduced in section
\ref{ssec-defs} ($m\I{c} = \lg M\I{c}$ etc.) reads:
\beq{eq-jmd}
l_* \!=\! 4.826 + 0.066z\I{i} - 0.01z\I{i}^2 + (3.186 - 0.129z\I{i} -
3.48m\I{c})m\I{c},
\eeq
and which reproduces the data of this work for $\Mc \ga 0.26$ so
perfectly that there is no need to establish a new relation. Together
with Eq. (\ref{eq-dMcdt}) below, Eq. (\ref{eq-jmd}) can be directly
integrated to obtain $L_*(\tau)$. For the purposes of synthetic AGB
evolution, the core mass at the instant of time when the helium flash
occurs can be approximated by
\beqa \label{eq-mcchef}
M\I{c, RGB tip} &=& (0.454 - 0.023z\I{i} + 0.059\Mi) \times \\ \nonumber
 & &	\left( 1 - (0.033 - 0.008z\I{i}) \e^{\Mi} \right),
\eeqa
which has a maximum of $M\I{c, RGB tip}$ approximately at an initial
mass of
\beq{eq-mcimax}
M\I{i, max} = - \ln (0.0169 - 0.005z\I{i}) - 2.935 - 0.022z\I{i}.
\eeq
For $\Mi \!<\! M\I{i, max}$, $M\I{c, RGB tip} \!=\! M\I{c, RGB tip}(M\I{i,
max})$ must be used.

When $M\I{c, RGB tip} \la 0.35$ according to Eq. (\ref{eq-mcchef}),
helium ignites under non-degenerate conditions, and there is no real
RGB. This is the case for $\Mi \ga 2.5$. For $\Mi \le 1.5$, there is
no systematic difference between OPAL and LAOL calculations ($\Delta
\Mc \le 0.005$). The transition from degenerate to non-degenerate
ignition turns out to be relatively model dependent. However, this
does not affect the AGB, since for $\Mi \ga 2$ the RGB tip luminosity
does not exceed $1500\Lsun$, and the transition timescales from
degenerate to non-degenerate are relatively short, so that almost no
mass is lost.

The following core helium burning phase does not directly influence
the AGB, since the luminosities are too low to drive significant mass
loss, except for blue horizontal branch stars with an already very
small mantle mass. These may directly evolve upward in the HRD and
form AGB manqu\'{e}e stars, which are not further considered here.

\subsection{The E-AGB}
\label{ssec-EAGB}

After core helium exhaustion, the helium shell initially burns
quiescently. The first TP occurs when the He-exhausted core approaches
the H-shell, at $M\I{H} - M\I{He} \la 10^{-2}$. $L_*$ rises from the
relatively low values typical for convective core helium burning up to
the high AGB luminosities. Since the evolutionary time scales on the
E-AGB are long compared to TP recurrence times, and since for stars
with $\Mi \ga 4$ the maximum luminosity attained on the E-AGB exceeds
the luminosity at the first TP, a considerable amount of mass may be lost
already on the E-AGB. This is interesting with respect to the
initial-final mass relation, since for $\Mi \approx 4$ there are
indications that the final mass is less or comparable to the core mass
at the first full thermal pulse (VW93).

On the E-AGB, there are competing contributions from H- and
He-burning, depending on $\Mi$ and the evolutionary phase, so that it
makes no sense to use a CMLR. Instead, the normalized E-AGB luminosity
$l\I{E}$ is described directly as a function of the E-AGB phase
$\phi\I{E} \equiv (\tau - \tau_0) / (\tau_1 - \tau_0)$, $0 \le
\phi\I{E} \le 1$:
\beq{eq-EAGB}
l\I{E} \;\equiv\; \frac{l_* - l_0}{l_1 - l_0} \;=\;
	v \phi\I{E} \,+\, (1 - v) \phi\I{E}^\beta.
\eeq
%
The mass of the helium exhausted core at the beginning of the E-AGB,
$M\I{He,0}(\Mi, \Zi)$, is calculated from $M\I{He,0} = {\rm max}
\left(q_1 +q_2 \times M\I{i}, q_3 +q_4 \times M\I{i} \right) $, where
the coefficients $q$ are listed in Table 1, for pop I and II, and can
be linearly interpolated in $\lg \Zi$. In case of ``breathing
pulses'', $M\I{He,0}$ may be larger by up to $(0.11 - 0.01\Mi)$.

Once $M\I{He,0}$ is known, the duration of the E-AGB, $\tau_1 -
\tau_0$, and the two parameters $v$ and $\beta$ in Eq. (\ref{eq-EAGB})
are approximately piecewise linear functions of $M\I{He,0}$ alone,
independent of $\Zi$. We find that $\lg (\tau_1 - \tau_0) \!=\!  {\rm
max} \left( (8.4 - 5.0 \times M\I{He,0}), (7.6 - 3.1 \times M\I{He,0})
\right) $,
$v \!=\! 
{\rm max} \left( 0.19, (-0.23 + 1.35 \times M\I{He,0}) \right)$ 
and $\beta 
\!=\! {\rm max} \left( (16.3 - 34.1 \times M\I{He,0}), 5.7 \right)$, 
for $M\I{He,0}$ varying between $0.14$ and $0.56$.


\begin{table}
\begin{center}
\begin{tabular}{lrrrr} \hline \hline
Pop & $q_1$ &\q $q_2$ &\q $q_3$ & $q_4$  \\ \hline
I   &0.168 &\q 0.0174 &\q --0.175 &0.125   \\
II  &0.150 &\q 0.0000 &\q   0.050 &0.120  \\
\hline \hline
\end{tabular}
\end{center}
\caption{The coefficients to calculate  $M\I{He,0}$ for $\Zi = 0.02$
and $10^{-4}$.}
\end{table}

\section{The luminosity on the TP-AGB as a function of time}
\label{sec-ltAGB}

\subsection{The core mass - luminosity relation (CMLR)}
\label{ssec-LA} 

\begin{figure}
 \centerline{\psfig{figure=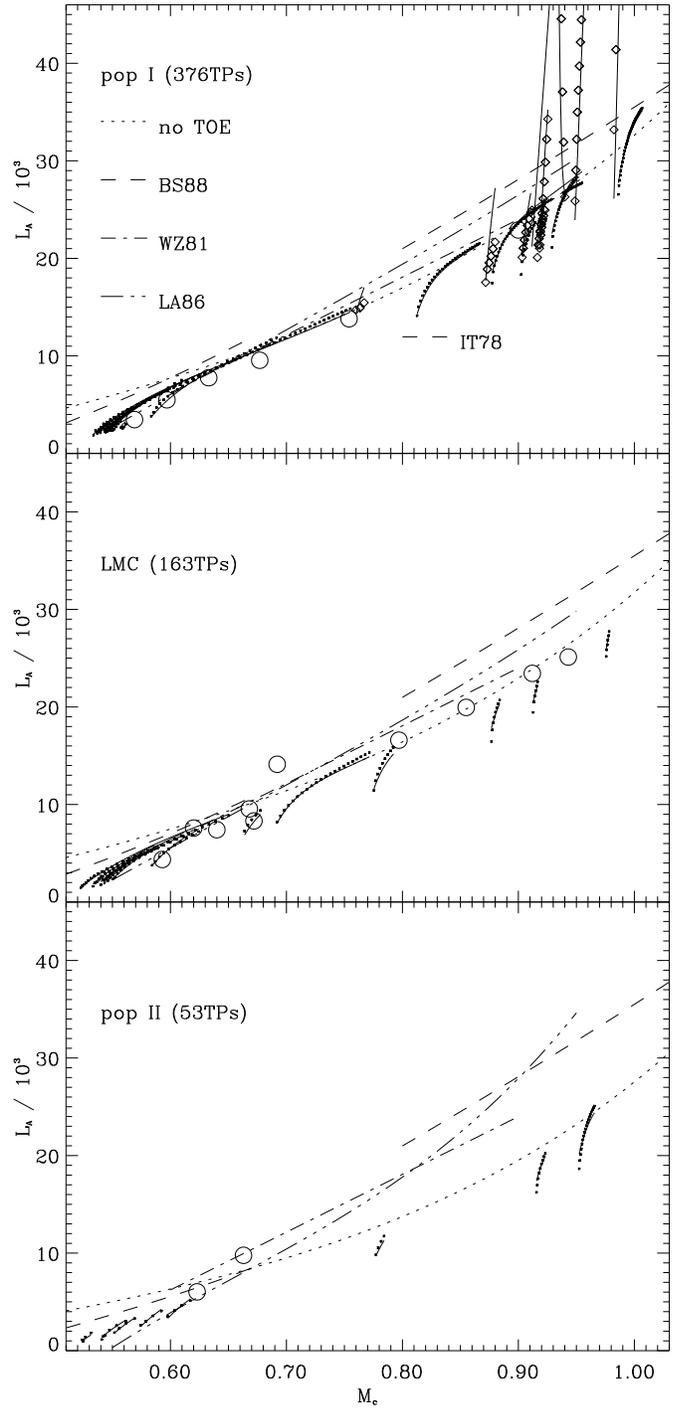,width=8.7cm}}
\caption[]{The maximum luminosity during quiescent hydrogen burning as
a function of the core mass and metallicity. {\it Dots}: data points
extracted from the calculations with $\alpha\I{MLT} = 1.5$;
$\Diamond$: data for $\alpha\I{MLT} > 1.5$ from the calculation with
$\Mi = 5$, DAM96 and B95. {\it Full lines} show the fit to the data
given by Eq. (\protect\ref{eq-LA}). $\bigcirc$: independent data, {\it not}
used for the fit, from Vassiliadis \& Wood (\cite{vw94}) for post-AGB
nuclei, i. e. objects with a vanishing mantle mass. The fit for this
case, and without TOE, is shown by the {\it dotted lines}. The {\it
other lines} represent CMLRs by various authors. For the formulae by
Boothroyd \& Sackmann (\cite{boo:sac}, BS88) and Lattanzio
(\cite{laz86}, LA86) the dependence on $\mu$, $Z\I{CNO}$ and $Y$,
respectively, has been taken into account approximately.  }
\label{fig-LA}
\end{figure}

One of the most important relations that come out of evolutionary
calculations is the maximum luminosity during
quiescent H-burning, here referred to as \LA, for which the following
expression according to Sect. 2.3 is derived:

%
\bsubeqa{eq-LA}{eq-LAa}
\LA \!=\! & & (18\,160 + 3980 z\I{i})(\Mc - 0.4468)    \\
\nsubeqn{eq-LAb1}
& + & 10^{2.705 \,+\, 1.649 \Mc} \times \\
\nsubeqn{eq-LAb2}
&   & \times \left(10^{0.0237 (\alpha - 1.447)
		\Mci^2 \Mm^2 (1 - \e^{ - \DMc / 0.01})}\right)  \\
\nsubeqn{eq-LAc}
& - & 10^{3.529 \,-\, (M\I{c,0} - 0.4468) \DMc / 0.01 }.
\esubeqa
The first term is recognized as the usual linear CMLR, but the slope
is flatter than usual. The term (\ref{eq-LAb1}) provides a correction:
when added to (\ref{eq-LAa}), with (\ref{eq-LAb2}) set to unity and
ignoring (\ref{eq-LAc}), the sum yields for $\Mc \ga 0.6$ the same
numerical values for \LA\ as most CMLRs that were fitted for low core
masses. On the other hand, for $\Mc \ga 0.95$ this sum approaches the
formula given by IT, which is applicable for massive cores only
(dotted lines in Fig. \ref{fig-LA})

The term (\ref{eq-LAb2}) corrects for HBB, which is character\-ized by
a steep initial increase of \LA, and a drop when the mantle mass $\Mm$
is reduced (B95, their figure 9). The calculations by DAM96 for $\Mi <
5$ show that HBB operates only for massive cores and mantles.  
The factor $(1 - \e^{-\DMc/0.01})$ mimics the initial increase of
$L_*$, which levels out after about the first ten TPs.  HBB strongly
depends on the MLT parameter, with the results of DAM96 that made use
of the CMT being described by $\alpha\I{MLT} = 2.75$. The data given
by VW93 (their figure 12), not used in Eq. (\ref{eq-LA}), are
remarkably well reproduced by the choice $\alpha = 2.25$.  


The last term, (\ref{eq-LAc}), corrects for the TOE, which is quite
important for $\Mc \la 0.65$, and independent of $Z$. 
Figure \ref{fig-LA} clearly shows that the CMLRs by WZ81
and LA86 are influenced by the TOE for low core masses.

\subsection{The core growth equation}
\label{ssec-dMcdt}

The equation that describes the core growth reads:
\beq{eq-dMcdt}
\frac{\D M_c}{\D t} = \frac{q}{X\I{m}}L\I{H},
\eeq
where $q$, the mass burnt per unit energy release, turns out to be
slightly metallicity dependent due to the contribution from the
pp-cycle for low $Z$. The calculations with a nuclear network yield
the mean value $q = ( (1.02 \pm 0.04) + 0.017 z)10^{-11}$ ($\Msun /
\Lsun$yr). The luminosity produced by H-burning that enters
Eq. (\ref{eq-dMcdt}) is smaller than the total luminosity, since
gravitational energy release due to the core shrinking and the flow of
burnt mantle material down to the core, together with relic helium
burning, additionally contribute during quiescent H-burning:
\beq{eq-LH}
\lg (L\I{H,A}/ \LA) = -0.012 - 10^{-1.25 - 113\DMc} - 0.0016\Mm.
\eeq
This equation was derived for the standard value of the MLT parameter,
which implies that the value of \LA\ to be used in Eq. (7) should be
calculated from Eq. (5), with the term (5c) put to unity as the
standard models did {\it not} experience any HBB.

\subsection{Further relations for luminosity extrema}
\label{ssec-LBCD}

The first observable effect of a TP is that the layers above the
helium shell source expand, thus the hydrogen shell is extinguished
and the total luminosity drops. This {\it ``rapid dip''} is the more
pronounced the smaller the mantle mass is, since the inner layers of
the mantle act as a reservoir of thermal energy that partly
compensates the initial fast luminosity drop. The rapid dip is utterly
unimportant for the secular evolution due to its very short duration,
but during this phase $|\D L_* / \D t|$ reaches the highest values a
single low-mass star can achieve. A fraction of roughly $10^{-3}$ of
all pulsating AGB stars are expected to show corresponding rapid
period changes. The fit formula given below 
strongly depends on $\Mm$ even for low
core masses:
%
\bsubeqa{eq-LB}{eq-LBa}
\LB & = & \left( 4.81(\Mc - 0.4865) + \Mm^{0.393} \right) \times \\
\nsubeqn{eq-LBb}
& & \left( 10^{2.879 \,+\, \Mc} -
10^{ 3.287 \,-\, (\Mci - 0.4865)\DMc / 0.01} \right).
\esubeqa

\begin{figure}[t]
 \centerline{\psfig{figure=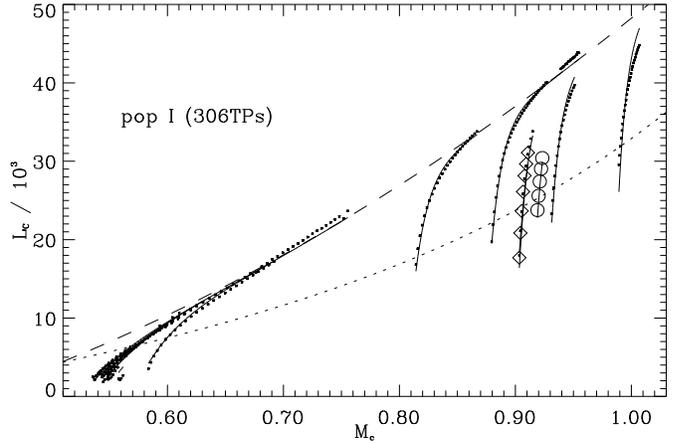,width=8.7cm}}
\caption[]{The luminosity peak after a TP as a function of the core
mass for $\Zi \!=\! 0.02$. {\it Dots}: data points from the calculations
with $\alpha\I{MLT}\!=\!1.5$ and used for the fit; $\Diamond$:
$\alpha\I{MLT}\!=\!2$, $\bigcirc$: data from DAM96, both not used for
the fit, which is shown by {\it full lines}. The appropriate values
for $\Mci$ are used for the individual sequences.
The {\it dotted line} depicts $\LA$ without TOE and HBB as in
Fig. \ref{fig-LA}. The {\it dashed line} is the asymptotic
relation Eq. (\ref{eq-LCa}). }
\label{fig-LC}
\end{figure}

\noindent
After the rapid dip, for $\Mc \la 0.7$ the layers above the inactive
H-shell start to contract, therefore releasing gravitational energy;
for heavier cores, $L_*$ simply follows $L\I{He}$ after a short
thermal adjustation phase. In either case a {\it ``rapid peak''}
emerges, in which the maximum total luminosity exceeds the quiescent
H-burning luminosity prior to the TP, except for the very first weak
TPs or in the case of strong HBB. Note that in many earlier
calculations these peaks were suppressed for large mantle masses due
to an insufficient numerical resolution. It turns out that a HBB term
does not improve the fit, and is therefore left out:
%
\bsubeqa{eq-LC}{eq-LCa}
\LC & = & 59~200\,M_c^2 - 10~950 - \\
\nsubeqn{eq-LCb}
& & 10^{2.559 \,+\, 1.951\Mci \,+\,
	 (44.7 \,+\, (2254\Zi - 147.6)\Mci)\DMc}
\esubeqa
Here Eq. (\ref{eq-LCb}) accounts for the TOE, which vanishes for $\DMc
\ga 0.03$.  In the asymptotic regime, the data nicely follow a
quadratic relation (Fig. \ref{fig-LC}). $\LC$ is an interesting
quantity for two reasons. First, the rapid peak is the only mechanism
capable of populating a high luminosity tail in the luminosity
function of a sufficiently large sample with $\Mi \la 3$. Second, if
it should be true that an avalanche mass loss starts as soon as some
luminosity threshold is exceeded (Tuchman et al. \cite{ytsb}), this would take place for the first time during a rapid
peak, unless HBB is operating so effectively that $\LA > \LC$. It is a
question of the interplay between the mass loss time scale and the
duration of the peak, if such a gasping mass loss episode influences
the secular evolution. 

\begin{figure}[t]
 \centerline{\psfig{figure=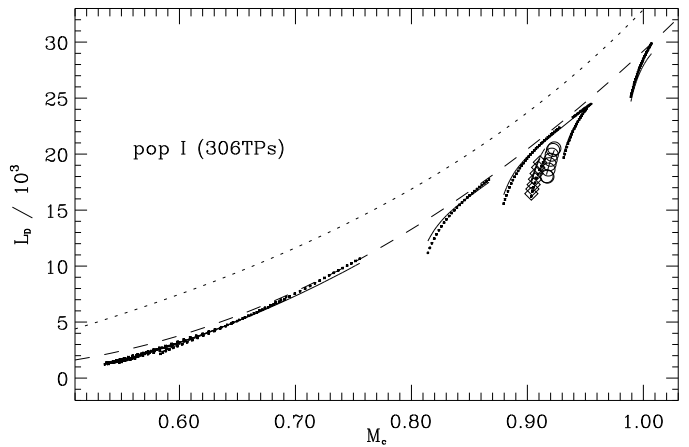,width=8.7cm}}
\vspace{-8mm}
\caption[]{The minimum luminosity at the beginning of a TPC as a
function of $\Mc$ for $\Zi = 0.02$. {\it Dashed line}: Eq. (\ref{eq-LDa}),
other symbols: see Fig. \ref{fig-LC}. }
\label{fig-LD}
\end{figure}

Finally, the helium burning declines and the H-shell recovers, giving
rise to an extended and long-lasting {\it ``slow dip''}. During the
quiescent part of a TPC, $\LD < L_* < \LA$. The fit, 
like Eqs. (\ref{eq-LB}) and (\ref{eq-LC}), needs no HBB correction:
\bsubeqa{eq-LD}{eq-LDa}
L\I{D} & = & (76360 + 6460 z\I{i}) (\Mc - 0.3881)^2 - \\
\nsubeqn{eq-LDb}
&& 10^{1.55 \,+\, 2.11\Mci \,+\, (59.7 -119.8\Mci)\DMc}.
\esubeqa

\subsection{The core mass-interpulse time relation}
\label{ssec-tip}

\begin{figure}
 \centerline{\psfig{figure=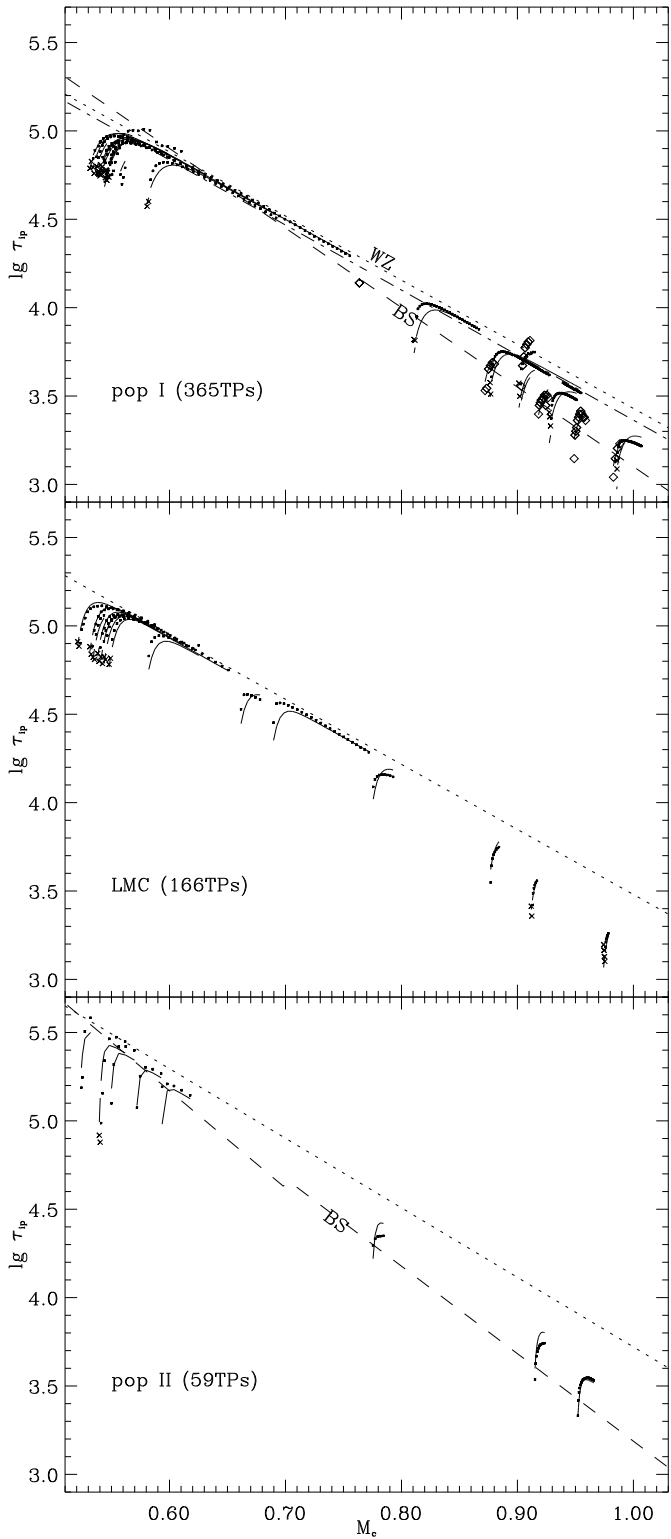,width=8.7cm}}
\vspace{-12mm}
\caption[]{The interpulse time as a function of the core mass and
metallicity. Data points taken from the calculations, used for the
fit, are shown as {\it dots}\/, the values given by Eq. (\ref{eq-tip})
as {\it full lines}. $\diamond$: data from DAM96, and from the run with
$\alpha\I{MLT} = 2$, $\Mi = 5$ ($\Mc \ga 0.9$).  $\times$: recurrence
time of pre-pulses, not used for the fit. {\it Dashed} and {\it
dash-dotted} lines: relations from BS88 and WZ81, respectively. {\it
Dotted} lines: the asymptotic relation Eq. (\ref{eq-tipa}). }
\label{fig-tip}
\end{figure}


The second most important quantity to come out of evolutionary calculations is
the core mass-interpulse time relation. The following relation is found:
 \bsubeqa{eq-tip}{eq-tipa}
\lg \tau\I{ip} & = & (-3.628 + 0.1337 z\I{i})(\Mc - 1.9454) -    \\
\nsubeqn{eq-tipb}
& & 10^{-2.080 \,-\, 0.353 z\I{i} + 0.200 (\Mm + \alpha - 1.5)} - \\
\nsubeqn{eq-tipc}
& & 10^{-0.626 \,-\, 70.30(\Mci - z\I{i}) \DMc}.
\esubeqa
Again, Eqs. (\ref{eq-tipb}) and (\ref{eq-tipc}) account for HBB and
the TOE, resp.  Interestingly, Eq. (\ref{eq-tip}) predicts the
recurrence time of pre-pulses with $\Mc < \Mci$ (crosses in
Fig. \ref{fig-tip}). For $\Mc \la 0.7$ and in the asymptotic regime,
the results of the present work agree well with relations for $\tip$
given by BS88 and WZ81. However, the latter disagree for $\Mc \!>\!
0.8$, and give lower and upper limits only, respectively.

Compared to the asymptotic exponential relation, the TOE reduces
$\tip$ by almost a factor of two. The influence of $\Zi$ on $\tip$ is
rather strong, in the sense that the interpulse times increase with
decreasing metallicity. This also means that the He-shell accretes
material processed by the H-shell for a longer time, so that the TPs
become more violent. The reason is that the plasma in a typical pop II
He-shell is more degenerate 
than in a pop I-shell. 
An important
consequence is that the third dredge-up occurs the earlier, i.e. for
lower core masses, the lower the metallicity.



A further quantity needed to construct secular lightcurves is the time
$\tau\I{D}$ that elapses between a TP and the slow dip, i.e. when
$L_* = \LD$. We find 
\beq{eq-dtd}
\lg (\tau\I{D} / \tip) = -1.01 + 0.20\, z\I{i} \,(\Mc - 1).
\eeq

\subsection{The core mass at the first thermal pulse}
\label{ssec-Mci}

\begin{figure}
 \centerline{\psfig{figure=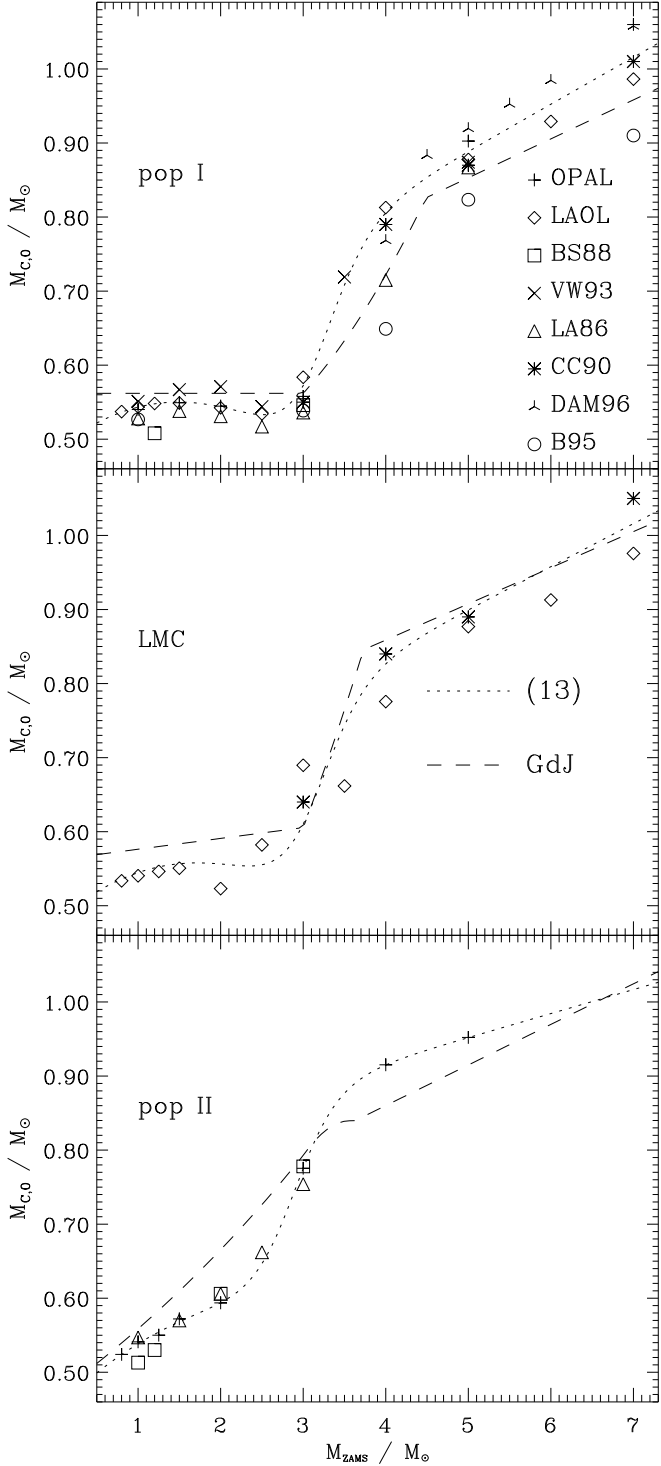,width=8.99cm}}
\caption[]{The core mass at the first TP as a function of the initial
mass and metallicity. The symbols show data from various authors
(CC90: Castellani et al. 1990). The lines show the
relation \ref{eq-mcia} (dots) and the one used by GdJ (dashed). }
\label{fig-Mc0}
\end{figure}

\begin{table}[b]
\begin{center}
\begin{tabular}{lrrrrrrrrrr} \hline \hline
Pop & $p_1$ &\q $p_2$ &\q $p_3$ & $p_4$ &\q $p_5$ & $p_6$ &\q $p_7$  \\ \hline
I   &0.0294 &\q 1.478 &\q 0.550 &0.0634 &\q 0.572 & 3.193 &\q 0.260  \\
II  &0.0213 &\q 2.589 &\q 0.592 &0.0324 &\q 0.790 & 2.867 &\q 0.260  \\
\hline \hline
\end{tabular}
\end{center}
\caption{The coefficients of Eq. (\protect\ref{eq-mci}) for $\Zi = 0.02$
and $10^{-4}$.}
\label{tab-mci}
\end{table}

\noindent
There is one quantity of outstanding importance left to be discussed,
the core mass at the first TP, i. e. $\Mci(\Mi, \Zi)$. It determines
the minimum remnant mass, and is therefore closely connected to
the initial-final mass relation. For $\Mi \la 2$, the stars develop
very similar degenerate He-cores (see section \ref{ssec-rgb}), and all
end up with $\Mci \approx 0.55$. On the other hand, massive stars with
$\Mi \ga 4.5$ experience the second dredge-up, which shifts the
H-He-discontinuity inward (in mass), with the resulting $\Mci$ being a
linear function of $\Mi$. In between, there is a transition region,
which is rather strongly influenced by computational details. For
stars in this transition region, $\D \Mci / \D \Mi \ga 0.25$, and
since $L_*$ increases strongly with $\Mc$, and the mass loss even much
more steeply with $L_*$, such transition stars, despite their larger
mass reservoir, have a {\it shorter} AGB lifetime than slightly less
massive counterparts. In WW94 the following parameterization was
introduced:

\bsubeqa{eq-mci}{eq-mcia}
\Mci(\Mi) & = &  \left( - p_1 (M\I{i} - p_2)^2 + p_3 \right) f + \\
\nsubeqn{eq-mcib}
	  &   &  \left( p_4 M\I{i} + p_5 \right) (1 - f),
			\;\;\;\;\;\;\;\;\; {\mathrm where} \\
\nsubeqn{eq-mcic}\;\;\;
f(M\I{i}) & = &  \left( 1 + \e^{\frac{M\I{i} - p_6}{p_7}} \right)^{-1}.
\esubeqa
Equation (\ref{eq-mcia}) mimics $\Mci$ for stars with $\Mi < 2.5$ that
experience the core helium flash, which is almost
constant. Eq. (\ref{eq-mcib}) accounts for the linear ingress of the
second dredge-up for $\Mi \ga 4.5$, and Eq. (\ref{eq-mcic}) mediates. All
parameters can be interpolated linearly in $\lg \Zi$, hence in Table
\ref{tab-mci} their numerical values are given for two standard
metallicities only.

A compilation of data from various authors is shown in
Fig. \ref{fig-Mc0}.  There is a tendency that more recent opacities
yield higher values of $\Mci$. Bl\"{o}cker's (B95) data points lie
systematically below all other's.
Obviously there is no systematic effect due to the inclusion of
semiconvection (LA86, DAM96). The compilation by GdJ systematically
deviates from the data for the LMC metallicity in the sense that for
low-mass stars it predicts too large values of $\Mci$.


%
%
\subsection{Luminosity variations during the thermal pulse cycle}
\label{ssec-lv}

In previous synthetic AGB calculations, the luminosity variations
during a thermal pulse cycle were either neglected (Renzini \& Voli
1981), or described by simple stepfunctions (GdJ, Marigo et
al. 1996). This is clearly insufficient if accurate theoretical
luminosity functions are to be predicted. This section completes the
description of $L_*(\tau)$ on the AGB.


At the instant of time when $L_* = \LD$, i. e. in the slow dip,
the H-shell is recovering from the preceding TP, and $L\I{H} \approx
L\I{He}$. It turns out that the luminosity produced by H-burning that
enters Eq. (\ref{eq-dMcdt}), when normalized to the value predicted by the
CMLR Eq. (\ref{eq-LA}) and Eq. (\ref{eq-LH}), has an almost universal form
(see also WZ81), independent of $\Zi$. Let $\phi \equiv \tau / \tip$
($\phi = 0$ at the TP initiating the TPC, when $\Mc = M\I{c,TP}$).
Then we define
\beq{eq-hlh}
h\I{H}(\phi) \equiv
	\frac{L\I{H}(\phi)}{L\I{H,CMLR}(\Mc(\phi), \Mci, M_*)} = 
	\frac{1 - \e^{-\beta\I{H} \phi^2}}{1 - \e^{-\beta\I{H}}},
\eeq
where $\beta\I{H} = 93.11 (M\I{c,TP} - 0.4569)$. There are small
systematic deviations for the first TP of each sequence,
which, however, have no influence 
on the integration of Eq. (\ref{eq-dMcdt}).

$L_*$ depends similarly on $\phi$. In the following, the arguments of
quantities given by a CMLR are omitted; the index ``TP'' denotes a
quantity as given by the corresponding CMLR for $\Mc = M\I{c,TP}$. The
form function $h\I{L}$ is given by
\beq{eq-hla}
h\I{L}(\phi) \equiv
	\frac{L_*(\phi) \frac{L\I{A,TP}}{L\I{A}(\Mc(\phi))} - L\I{D,TP}}
	     {L\I{A,TP} - L\I{D,TP}} =
	\frac{1 - \e^{- \beta\I{L} \phi_*^{b\I{L}}}}
	     {1 - \e^{- \beta\I{L}}},
\eeq
where $\beta\I{L} \!=\! 7.95$ and $b\I{L} \!=\! 2.13\, (1.331 \!-\!
M\I{c,TP})$, and $\phi_* \!\equiv\! \max \{ \phi \!-\! 0.1, 0 \}$
accounts for the behaviour shortly after the slow dip. Note that
$h\I{L}(1) \!>\! 0.999 \!\approx\!  1$, and that $0.1 \!\la\! \phi
\!<\! 1$ due to Eq. (\ref{eq-dtd}).





Let now $\tilde{\phi} \equiv (\tau - \tau\I{C}) / \tip(\Mc, \Mci,
M_*)$ be a shifted TPC phase, so that $\tilde{\phi} = 0$ at the rapid
peak when $L_* \!=\! \LC$. The peak itself can be approximated by a
parabola, followed by an expo\-nential decline. 
The luminosity from the rapid peak until $L_* = \LD$ can
well be described by:
\beqa \label{eq-rpeak}
l_*(\tilde{\phi}) & = & \lg \LC -  \\ \nonumber
& & \left\{ \begin{array}{ll}
a \tilde{\phi}^2, &  \mbox{if $\tilde{\phi} \le \frac{b}{2 a}$,} \\
b \tilde{\phi} - \frac{b^2}{4 a},
	&  \mbox{if $\tilde{\phi} \le \frac{b}{4 a} + \frac{\Delta\I{N}}{b}$,} \\
\Delta\I{N} ( 1 - f\I{N}) - f\I{N} \left(- b \tilde{\phi} + \frac{b^2}{4 a} \right)
   &  \mbox{else.}
	\end{array}
\right.
\eeqa
The physical meaning of $\Delta\I{N}$ and $f\I{N}$ is, that when the
luminosity $l_* = \lg \LC - \Delta\I{N}$ is reached, the time scale
for the decline changes from $\tip/b$ to $\tip/(f\I{N} b)$. The
parameters $\lg a, \lg b$, $\lg \Delta\I{N}$ and $\lg f\I{N}$ are
functions of $\Zi$, $\Mc$ and $\Mci$ and can again be described by
linear relations with a TOE correction for $\lg a$ and $\lg b$. They
obey the simplified relations
\beqa
 \lg a 		 & = &  7.23 - 4.40\, M\I{c}, \\
 \lg b 		 & = &  3.30 - 2.91\, M\I{c}, \\
 \lg \Delta\I{N} & = &  0.08 - 1.50\, M\I{c}, \mbox{\hspace{0.5cm}and} \\
 \lg f\I{N} 	 & = & -1.84 + 1.79\, M\I{c}.
\eeqa
Note that the duration of the rapid peak does not scale like
$\tip$. Relative to $\tip$, it lasts the longer the more massive the
core is, i. e. the shorter $\tip$. 

\section{Properties of the mantle}
\label{sec-mntl}

\subsection{The method}
\label{ssec-introm}

Several physical aspects of AGB evolution depend on properties that
are almost exclusively determined by the mantle. The justification is
that as long as there is a radiative transition region between the
core and the lower boundary of the mantle convection zone, the
influence of the two outer boundary conditions on the core is
negligible (WZ81). On the other hand, since the luminosity is
essentially constant, the core, which prescribes $L$, can to first
order be considered to be just a gravitating point source for the mantle. 

Now consider $L_*, M_*, \Mc$ and the chemical composition to be given.
The latter, owing to mixing during earlier evolutionary phases, is
homogeneous up to and including the upper parts of the H-shell. Two
outer boundary conditions for the mantle are given immediately: the
effective stellar surface radius $R_*$ is
given by the Stefan-Boltzmann law
and from some model for the optically thin outer envelope one can
derive $P_*(L_*, M_*, T_*)$. By assuming a value for $T_*$, the four
stellar structure equations for $L, T, P$ (or $\rho$) and $R$ as
functions of the Lagrangean mass coordinate $M\I{r}$ can now be
integrated inward. The transition to the core is well defined (see
also Wagenhuber \& Tuchman \cite{wag:yt}) by a very steep rise of $T$
and $P$ by one and four, resp., orders of magnitude or more. In this
way, a relation $\Mc(T_*)$ is defined that can be inverted numerically
to yield $T_*$ as a function of all other quantities, defining a three
dimensional manifold in the four dimensional parameter space $(T_*,
L_*, M_*, \Mc)$.  Full stellar evolution calculations of course
establish a connection between $L_*$ and $\Mc$. In order to study pure
mantle properties, however, it is much more convenient to treat them
as independent variables to get rid of the complicated time dependent
problem one is confronted with in full calculations.




In the following, a set of data relevant for HBB is discussed,
together with {\it consistent} effective temperatures, for various
model assumptions that made use of OPAL (see Sect. \ref{ssec-err})
and more recently published opacity tables.

\subsection{The effective temperature}
\label{ssec-teff}

\begin{figure*}[p]
 \centerline{\psfig{figure=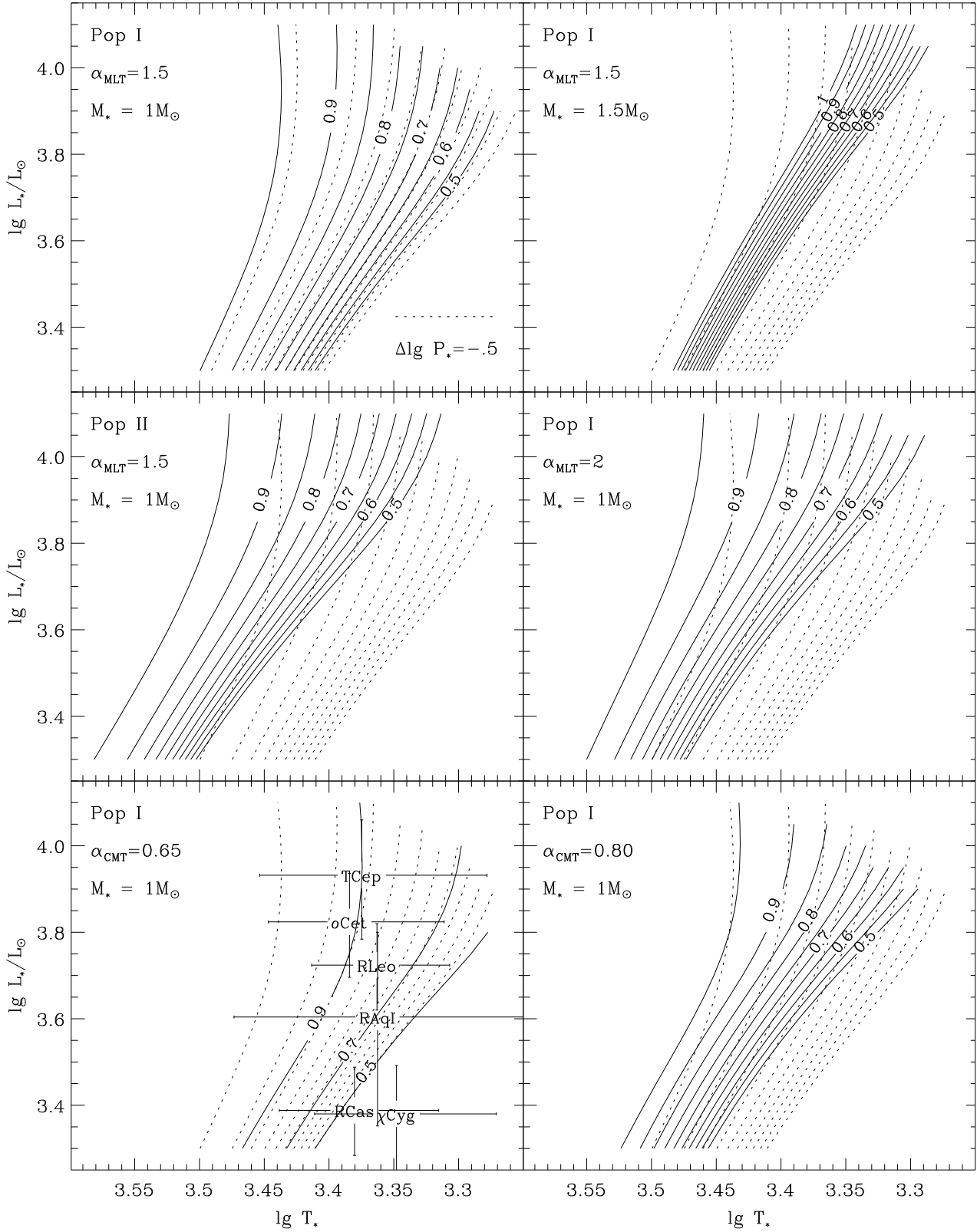,width=17cm}}
\vspace*{-.75cm}
\caption[]{The effective temperature as a function of $L_*, M_*, \Mc$
and $Z$, and the influence of various model assumptions by means of
the example $M_* = 1$. The standard case is depicted in the upper left
box; the full lines are $T_*(L_*)$ for $\Mc = 0.5, \ldots, 0.95\Msun$.
This case is repeated in all other boxes as dotted lines for
comparison. In the upper left box itself, the dotted lines are
obtained by reducing the pressure at the outer boundary condition by a
factor of three as compared to the usual relation $P_*(L_*, M_*,
T_*)$. In the lower left box, recent observational data (see text) are
overplotted with $1\sigma$ error boxes.}
\label{fig-teff}
\end{figure*}


Figure \ref{fig-teff} illustrates how $T_*(L_*, M_*, \Mc)$ depends on
the various variables and physical model assumptions. Let us first
discuss the standard case (pop~I, $M_* = 1$, MLT; upper left box). For
a given core mass, $t_*$ decreases almost linearly for increasing
$l_*$ as known from the Hayashi theory, the exact value of the slope
depending on the opacity as a function of $P$ and $T$, here $\D t_* /
\D l_* \approx -0.2$.  When $l_*$ approaches the Eddington luminosity
($L\I{Edd}$), $t_*$ becomes locally almost independent of $l_*$, and
there is even a bending towards hotter $t_*$ for large $\Mc$. Such
mantles are no longer fully convective. It is obvious that the
pressure at the photosphere has a minor influence (dotted lines, upper
left box), since on the AGB for $l_* > 3$, $p_* < 3$ for all
atmosphere models considered.

If the mantle mass is reduced, i. e. for growing $\Mc$ with $M_*$ kept
constant, $t_*$ {\it increases}. The influence of $\Mc$ on $t_*$ is
the less pronounced the larger $M_*$ is; for $M_* \ga 2$ it is already
almost negligible. Increasing $M_*$, while $\Mc =$ const, leads to
higher effective temperatures and vice versa without changing $\D t_*
/ \D l_*$ significantly (upper right box). This effect is responsible
for the trend towards lower effective temperatures on the upper AGB as
the objects lose mass, as long as $\Mm \ga 0.2$. If $\Mm$ becomes
smaller, the former effect takes over, and for $\Mc \approx$ const a
shrinking total mass leads to {\it higher} $T_*$, and thus to the
departure of proto-PN objects from the AGB.

Increasing the MLT parameter has nearly the same effect as reducing
the metallicity (middle row): it shifts the whole set of curves
towards higher effective temperatures without changing the influence
of $\Mc$. It turns out that there is no qualitative difference between
the MLT and the Canuto-Mazzitelli theory (\cite{can:maz}, CMT, lower
row).  Quantitatively, there is no single value of $\alpha\I{CMT}$
which reproduces $t\I{*,MLT}$ for all $\Mc$ and $l_*$ to within
300K. This difference, however, is much smaller than the errors of the
available observational determinations of $t_*$. Recent data (van
Leeuwen \ea, \cite{HIPPARCOSb}) that make use of \mbox{\sc hipparcos}
parallaxes,
shown in the lower left box of Fig. \ref{fig-teff}, support the lower
$T_*$-scale. E. g. $\chi$ Cyg can well be explained to be a low-mass
object ($M_* \approx 0.75$) with a low-mass progenitor, so that $\Mci
\approx 0.535$, in one of its first TPCs ($\Mc = 0.540$).

\subsection{The bottom of the convective mantle}
\label{ssec-tb}

\begin{figure*}[p]
 \centerline{\psfig{figure=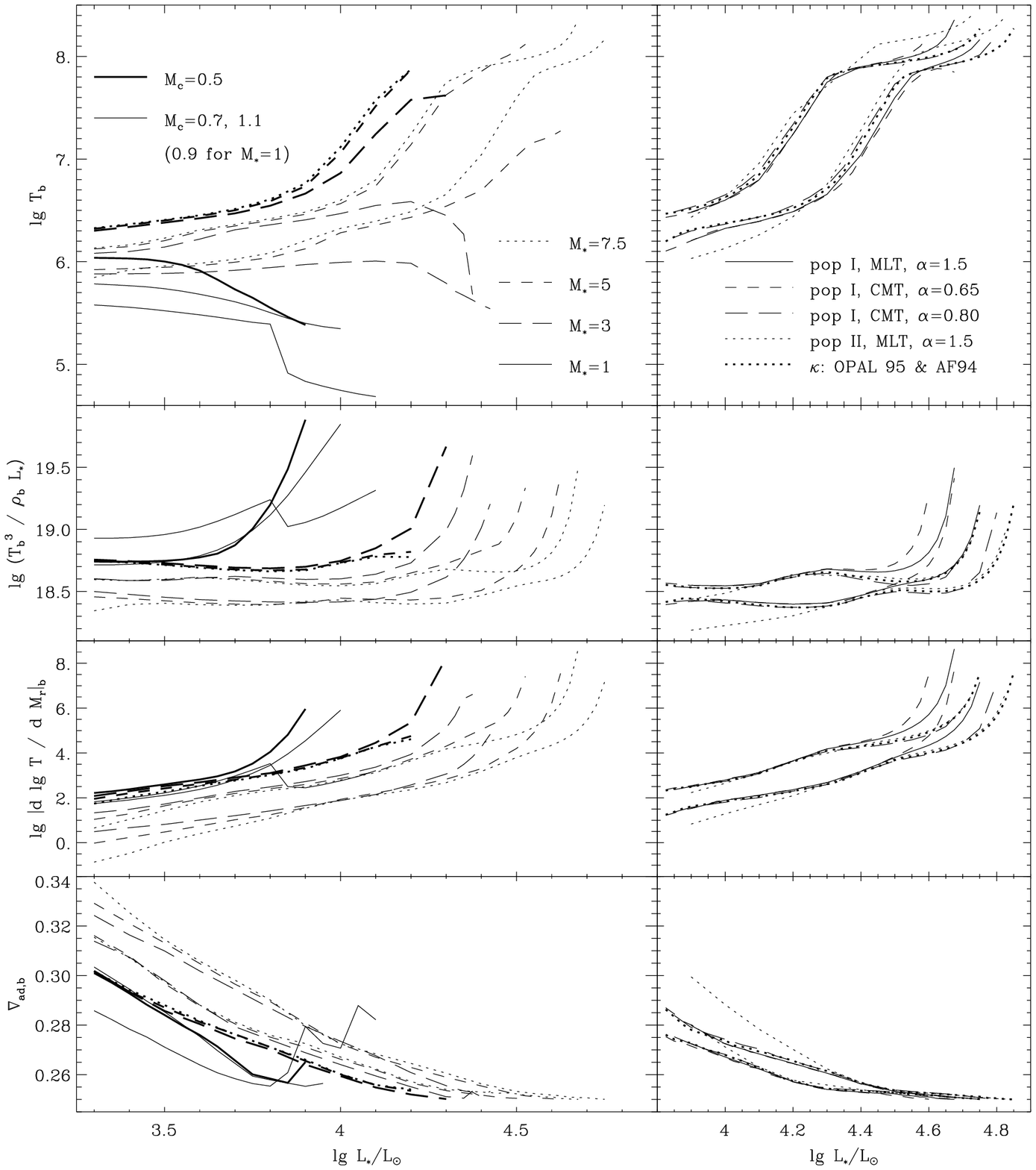,width=17.5cm}}
\caption[]{The temperature and some quantities characterizing the
overall conditions at the bottom of the convective mantle (see text)
as functions of $L_*, M_*$ and $\Mc$ for the standard case (pop~I,
MLT, $\alpha=1.5$, left column) and for various physical assumptions
(right column). Left: the thick lines show the data for
$\Mc = 0.5$ and different total masses as indicated. The thin lines
illustrate the effect of $\Mc$ (for $M_* = 1$, the highest core mass
is $0.9\Msun$, in all other cases $1.1\Msun$). Right: The same
quantities for $M_* = 7.5$ and $\Mc = 0.7$ and $1.1$ (left resp. upper
set of curves) for various physical model assumptions.  }
\label{fig-tb}
\end{figure*}

With respect to HBB, $T\I{b}(L_*, M_*, \Mc)$ together with $\rho\I{b}$
determines the rate of nuclear reactions (see e.g.  Scalo et al.  
\cite{scalo:d:u}). A fact that has been widely disregarded in
the past is that also the steepness of the $T$- and $P$-profiles is
decisive for HBB, since it, together with the mixing time scale,
determines the amount of matter subject to burning. There are two
extreme possibilities: On the one hand, the fuel could be burnt
locally at the bottom of the mantle convection zone. On the other
hand, if the mixing is sufficiently effective, the mantle material as
a whole at any instant of time could be subject to nuclear processing.
The local
mixing time scales, which are very
poorly known from the MLT, determine the actual situation between
these extremes. As long as the uncertainties due to the convection
formalism dominate to such an extent, it is well justified to use
parametrized descriptions instead of time consuming full stellar
evolution calculations.

The most important quantity is of course the temperature at the base
of the convective mantle (Fig. \ref{fig-tb}, upper left box).  The
present work qualitatively confirms the results obtained by Boothroyd
et al. (\cite{boo:s:a}) by means of full stellar evolution
calculations, and work out the basic dependencies more clearly, like
the shift of $T\I{b}$ caused by breathing pulses (their Figure 1c),
which can be easily explained as the effect of an increased core mass.

For sufficiently large total masses $M_* \ga 2$, $t\I{b}$ is roughly
independent of $M_*$.  Cool bases ($t\I{b} \la 7$) occur for $l_* \la
4$. If the luminosity increases beyond some critical value depending
on $\Mc$, $T\I{b}$ either rises approximately like $T\I{b} \propto
L_*^5$ for low core masses, or drops, since the inner parts of the
mantle become radiative for large $\Mc$ and small $M_*$ (in
Fig. \ref{fig-tb}, this is the case for $l_* > 4.2$, $M_* = 3$ and
$\Mc \ge 0.7$). At $t\I{b} \approx 7.9$, the rise flattens, unless
$l_*$ gets very close to the Eddington luminosity. For a given set of
opacity tables, ``hot bottoms'' can occur for sufficiently large
luminosities only if the ratio $\frac{\Mc}{M_*}$ is {\it less} than
some constant that depends solely on $Z$ and the convection
description. A useful approximation is 
\beq{eq-McMhot}
\frac{\Mc}{M_*} < \left\{ \begin{array}{ll} 0.2\,  \alpha\I{MLT} - 5Z, &
\mbox{\mathrm for MLT, and} \\ 0.5\, \alpha\I{CMT} - 5Z - 0.08, &
\mbox{\mathrm for CMT.}  \end{array} \right.  
\eeq
The influence of the physical models is only moderate
(Fig. \ref{fig-tb}, upper right box) and leads predominantly to a
shift of $t\I{b}(l_*)$ along the $l_*$-axis, which is equivalent to a
rescaling of $\Mc$. Every change of the physical model input that
leads to higher effective temperatures, like lower opacities at low
temperatures, lowers the threshold for $l_*$ above which HBB takes
place. Changes of the opacities $\kappa(T,P)$ can be compensated by
varying $\alpha$ appropriately, as can be clearly seen for the pair of
cases shown with long dashed and thick dotted lines.  For the former,
CMT and $\alpha\I{CMT} = 0.80$ was used, for the latter MLT, $\alpha =
1.5$ and recent OPAL opacities with low temperature extensions
according to Alexander \& Ferguson (\cite{alex:ferg}, labelled AF94 in
the figure). For the pop~II models, which imply a drastic reduction of
$\kappa$ at low temperatures, also the plateu value of $t\I{b}$ is
increased significantly (dotted lines in Fig. \ref{fig-tb}, upper
right box).

In order to calculate nuclear reaction rates, the density at the
convective base must be given, too. Since the mantle convection zones
are almost entirely adiabatic, $\rho\I{b}$ also fixes the entropy. For
a sufficiently large total mass, and unless $L_*$ approaches the
Eddington limit, the simple relation \beq{eq-t3rhol} \lg
\frac{T\I{b}^3}{\rho\I{b} L_*} = 18.9 - 0.4 \Mc \eeq
holds remarkably well over one order of magnitude of $L_*$
(Fig. \ref{fig-tb}, upper middle row) and may be used to estimate
$\rho\I{b}$.

As has been already mentioned, the quantity $\left. \frac{\D \lg T}{\D
M\I{r}} \right|\I{b}$ (Fig. \ref{fig-tb}, lower middle row) determines
the amount of matter subject to nuclear processing. It increases
approximately as $\propto L_*^3$, and even superexponentially when
$L_* \rightarrow L\I{Edd}$. This means, that the effect of reaching
very high base temperatures could be entirely compensated by the fact
that almost no fuel is left. It must be emphasized, that the nuclear
yields will depend extremely sensitively on the mixing time scales in
such a situation, which are not correctly predicted by the MLT!

In the lower row of the figure, the adiabatic gradient $\nabla\I{ad,b}
\equiv \left. \frac{\D \ln T}{\D \ln P} \right|\I{ad,b}$ is shown. Let
$h \equiv \left. \frac{\D \ln \rho}{\D \ln T} \right|\I{b}$, with $1.8
\la h \le 3$, denote the steepness of the density profile. Then
$\nabla\I{ad,b}$, the ratio $\beta$ of the gas pressure to the total
pressure, including that of radiation, and $h$ are (exactly) connected
by the relations
\beq{eq-nad}
\nabla\I{ad,b} = \frac{3}{4}\frac{2 h - 7}{2 h^2 - 6 h -3}
	= \frac{4 - 3 \beta}{16 - 12 \beta - \frac{3}{2} \beta^2}
\eeq
as long as degeneracy plays no role for the EOS. In any case,
$\rho\I{b} < 10$g cm$^{-3}$, so that this assumption is well fulfilled.
The uniform decrease of $\nabla\I{ad,b}$ as $L_*$ increases
demonstrates the increasing contribution of the radiation pressure to
$P\I{b}$. For $L_* \rightarrow L\I{Edd}$, $\beta \rightarrow 0$ and
$\nabla\I{ad,b} \rightarrow 0.25$. $t\I{b}(l_*)$ reaches the plateau
value of $t\I{b} \approx 7.9$ approximately when the radiation
pressure dominates, $\beta < 1/3$, or $\nabla\I{ad,b} < 0.255$.

%
%
%


\section{Summary}

We present analytical formulae that describe the evolution of a star
on the AGB. This outdates all previous work on this subject up to now
for several reasons. First, all formulae are based on a (nearly)
homogeneous set of up-to-date full stellar evolution models that cover
all relevant masses and a large range in metallicity. Second, the
analytical formulae take into account several important features found
in the full calculations, and largely neglected so far in previous
analytical descriptions: the secular variations of the luminosity
during a thermal pulse cycle, turn-on effects during the first few
pulses, and hot bottom burning. What these descriptions and formulae
can possibly not account for, is the influence or feed-back of
dredge-up (or nucleosynthesis in general) on the evolution of the
star. \\

\noindent
In future work we plan to include a description of the variations of
the various chemical species during 1st, 2nd and 3rd dredge-up and
HBB, which were not addressed in the full stellar evolution models. \\

\noindent
We are currently writing a numerical code to implement the analytical
formulae described in this paper. When the chemical description is
included we are in a position to improve upon the AGB population
synthesis models of GdJ or Marigo et al. (1996). With such a code we
are also in a position to investigate new data on AGB stars in
extragalactic systems that will become available with the current and
new generation of 8-10m telescopes.

\begin{acknowledgements}
JW is indebted 
to Dr. Achim Weiss for comments on the manuscript and his general effort
for  him at the MPA. MG acknowledges help from Henrik Spoon (MPE) in
understanding IDL at a very late stage in the preparation of this manuscript.
\end{acknowledgements}

%
%




\begin{thebibliography}{}

\bibitem[1994]{alex:ferg}
Alexander D.R., Ferguson J.W., 1994, ApJ 437, 879  (AF94)
\bibitem[1989]{an:gr}
Anders E., Grevesse N., 1989, Geochimica et Cosmochimica acta 53, 197
\bibitem[1969]{arnett}
Arnett W.D., 1969, Ap \& Sp. Sci. 5, 180  
\bibitem[1995]{bloe:a}
Bl\"ocker T., 1995, A\&A 297, 727 (B95)
\bibitem[1988]{boo:sac}
Boothroyd A.I., Sackmann I.-J., 1988, ApJ 328, 632 (BS88)
\bibitem[1993]{boo:s:a}
Boothroyd A.I., Sackmann I.-J., Ahern S.C., 1993, ApJ 416, 762  
\bibitem[1991]{can:maz}
Canuto V.M., Mazzitelli I., 1991, ApJ 370, 295  (CMT)
\bibitem[1996]{cas2:tor}
Cassisi S., Castellani V., Tornamb\`e A., 1996, ApJ 459, 298  
\bibitem[1985]{ccpt}
Castellani V., Chieffi A., Pulone L., Tornamb\`e A., 1985, ApJ 296, 204  
\bibitem[1990]{cc90}
Castellani V., Chieffi A., Straniero O., 1990, ApJS 74, 463  (CC90)
\bibitem[1996]{dan:maz}
D'Antona F.D., Mazzitelli I., 1996, ApJ 470, 1093  (DAM96)
\bibitem[1992]{flei:gau}
Fleischer A.J., Gauger A., Sedlmayr E., 1992, A\&A 266, 321  
\bibitem[1996]{frost}
Frost C., Lattanzio J., 1996, ApJ 473, 383
\bibitem[1993]{gro:dej}
Groenewegen M.A.T., de Jong T., 1993, A\&A 267, 410  (GdJ)
\bibitem[]{}
Groenewegen M.A.T., de Jong T., 1994, A\&A 283, 463 
\bibitem[]{}
Groenewegen M.A.T., van den Hoek L.B., de Jong T., 1995, A\&A 293, 381
\bibitem[]{}
Herwig F., Bl\"ocker T., Sch\"onberner D., El Eid E., 1997, A\&A 324, L81
\bibitem[1983]{iben:renz}
Iben I. Jr, Renzini A., 1983, Ann.~Rev.~A\&A 21, 271
\bibitem[1978]{ib:tr}
Iben I. Jr, Truran J.W., 1978, ApJ 220, 980  (IT78)
\bibitem[1997]{jim:mcd}
Jimenez R., MacDonald J., 1998, MNRAS, submitted
\bibitem[1991]{uffe}
J\o rgensen U.G., 1991, A\&A 246, 118
\bibitem[1986]{laz86}
Lattanzio J.C., 1986, ApJ 311, 708  (LA86)
\bibitem[1996]{marigo}
Marigo P., Bressan A., Chiosi C., 1996, A\&A 313, 545  
\bibitem[1986]{mazz:dant}
Mazzitelli I., D'Antona F., 1986, ApJ 308, 706
\bibitem[1996]{mow:jor}
Mowlavi N., Jorissen A., Arnould M., 1996, A\&A 311, 803  
\bibitem[1970]{pac:a}
Olofsson H., Carlstr\"{o}m U., Eriksson K. et al., 1990, A\&A 230, L13  
\bibitem[1990]{olo:ce}
Paczy\'{n}ski B., 1970, Acta Astr. 20, 47  
\bibitem[1975]{pac:b}
Paczy\'{n}ski B., 1975, ApJ 202, 558  
\bibitem[1975]{reimers}
Reimers D., 1975, Mem. Soc. Roy. Li\`{e}ge, 6, VIII, 369  
\bibitem[1981]{renz:voli}
Renzini A., Voli M., 1981, A\&A 94, 175
\bibitem[1992]{OPAL}
Rogers F.J., Iglesias C.A., 1992, ApJS 79, 507  
\bibitem[1989]{rus:bes}
Russell S.C., Bessell M.S., 1989, ApJS 70, 865  
\bibitem[1990]{rus:dop}
Russell S.C., Dopita M.A., 1990, ApJS 74, 93  
\bibitem[1975]{scalo:d:u}
Scalo J.M., Despain K.H., Ulrich R.K., 1975, ApJ 196, 805  
\bibitem[1979]{schb:a}
Sch\"onberner D., 1979, A\&A 79, 108
\bibitem[]{}
Staniero O., Chieffi A., Limongl M., et al., 1997, ApJ 478, 332 
\bibitem[1979]{ytsb}
Tuchman Y., Sack N., Barkat Z., 1979, ApJ 234, 217
\bibitem[1997]{HIPPARCOSb}
van Leeuwen F., Feast M.W., Whitelock P.A., Yudin B., 1997, MNRAS 287, 955
\bibitem[1993]{vw93}
Vassiliadis E., Wood P.R., 1993, ApJ 413, 641 (VW93)
\bibitem[1994]{vw94}
Vassiliadis E., Wood P.R., 1994, ApJS 92, 125
\bibitem[1996]{phd}
Wagenhuber J., 1996, PhD thesis, TU M\"unchen
\bibitem[1996]{wag:yt}
Wagenhuber J., Tuchman Y., 1996, A\&A 311, 509  
\bibitem[1994]{wag:wei:a}
Wagenhuber J., Weiss A., 1994, A\&A 286, 121 (WW94)
\bibitem[1990]{wei:trur}
Weiss A., Truran J.W., 1990, A\&A 238, 178  
\bibitem[1990]{WKM}
Weiss A., Keady J.J., Magee N.H., 1990, 
Atomic Data and Nuclear Data Tables 45, 209  
\bibitem[1996]{wwd}
Weiss A., Wagenhuber J., Denissenkov P.A., 1996, A\&A 313, 581
\bibitem[1986]{woo:fau}
Wood P.R., Faulkner, D.J., 1986, ApJ 307, 659  
\bibitem[1981]{wo:za}
Wood P.R., Zarro D.M., 1981, ApJ 247, 247  (WZ81)

\end{thebibliography}
\end{document}